\documentclass[10pt,a4paper]{book}
\usepackage{frontier04}
\begin{document}
\newcommand{\anta}{{\sc Antares}}
\newcounter{ctr}
\setcounter{ctr}{\thepage}
\addtocounter{ctr}{6}

\talktitle{{\sc Antares} Status \& Milestones : \newline
{\small News from Deep-Sea}}

\talkauthors{Thierry Pradier \structure{a}, 
             for the {\sc Antares} Collaboration\footnote{{\em http://antares.in2p3.fr}}}

\begin{center}
\authorstucture[a]{Université Louis-Pasteur  
                   \& Institut de Recherches Subatomiques - {\sc IReS},  \newline
                   23 rue du Loess BP 28 - F67037 Strasbourg, France}
		   \end{center}

\shorttitle{{\sc Antares} : Status \& Milestones} 

\firstauthor{Th. Pradier}

\begin{abstract}
The \anta~project aims to build a deep-sea Cherenkov Telescope for High Energy Neutrino Astronomy located in the Mediterranean Sea.
The experiment, currently in the construction phase, has recently achieved an important milestone : the operation of
a prototype line and of a line with monitoring instruments. These deployments allowed a 
thorough understanding of environmental parameters.

\end{abstract}

\section{High Energy Neutrinos and AstroParticle Physics}

The advantage of using neutrinos as new messengers lies on their weak interaction cross-section:
unlike protons or gammas, they provide a cosmolo-gical-range, unaltered information from the very heart of their sources.
The drawback is that their detection requires a huge detection volume.
In all those aspects, neutrinos are comparable to gravitational waves (GW).
This comparison applies also to the sources of high energy neutrinos themselves.

Such sources, either galactic or extra-galactic, are compact objects, involving
relativistic movements of masses and particles, all necessary ingredients to have an efficient gravitational
wave emission (see section \ref{microq}). Most of those sources have already been extensively observed in $\gamma$: supernovae and their remnants,
pulsars, microquasars and AGNs, gamma-ray bursters. The main question is to know whether or not those 
photons are produced by electrons {\it via} inverse compton/synchrotron emission
or by protons/nuclei {\it via} production of neutral and charged pions, which decay to produce photons and neutrinos.

\section{The \anta~Neutrino Telescope}

\anta~can be seen as a fixed target experiment: a cosmic neutrino
interacts in the Earth and produces a muon that propagates in sea water. The Cherenkov light emitted
by the muon is detected by an array of photomultipliers arranged in strings, able to reconstruct the energy and direction of
the incident muon/neutrino.

The main physical backgrounds are twoflod. Atmospheric muons, produced in the upper atmosphere
by the interaction of cosmic rays, can be discarded because of their downward direction. Atmospheric neutrinos on the
other hand are
more delicate to identify: produced on the other side of the Earth, they have exactly the same signature as the cosmic signal \anta~awaits for.
They also represent a powerful calibration tool.

\subsection{The \anta~Collaboration and the detector}

The goal of the European collaboration \anta~\cite{anta} is thus to build an underwater telescope dedicated to high energy
 neutrinos. Around 20 particle physics laboratories, astrophysics and oceanography institutes are taking part in the project.
The selected site is in the Mediterranean Sea, 40 km from Toulon (Southern France), at a depth of 2500m. 

\begin{figure}[h!]
\begin{center}
\epsfig{figure=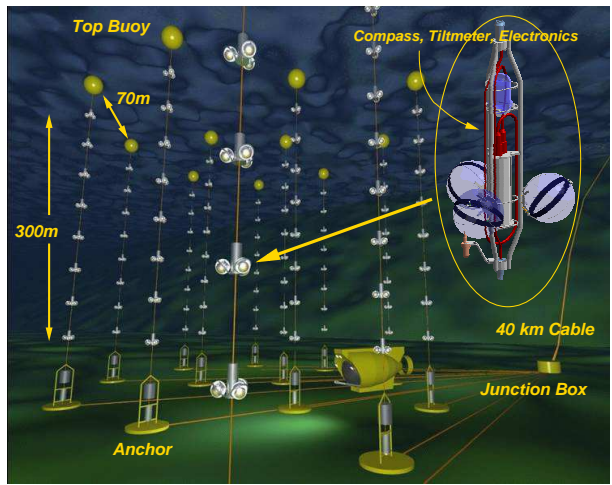,width=8.cm}
\end{center}
\caption{Artist's view of the final \anta~Neutrino Telescope.} 
\label{antares_layout}
\end{figure}

The detector consists in twelve lines, each one being composed of 25 storeys. On each storey, 3 photomultipliers are looking 
downward, to be sensitive to upward-going muons only. The layout of the detector is described in Figure \ref{antares_layout}.

\subsection{Physics Performance}

The performances of \anta~as a Neutrino Telescope are mainly estimated studying the effective are, as defined below 
(which convolved with a neutrino flux
determines the event rate) and background rejection capabilities, mostly based on the angular resolution 
and energy reconstruction.

\begin{figure}[t!]
\vskip -0.25cm

\centerline{\hskip -0cm \epsfig{figure=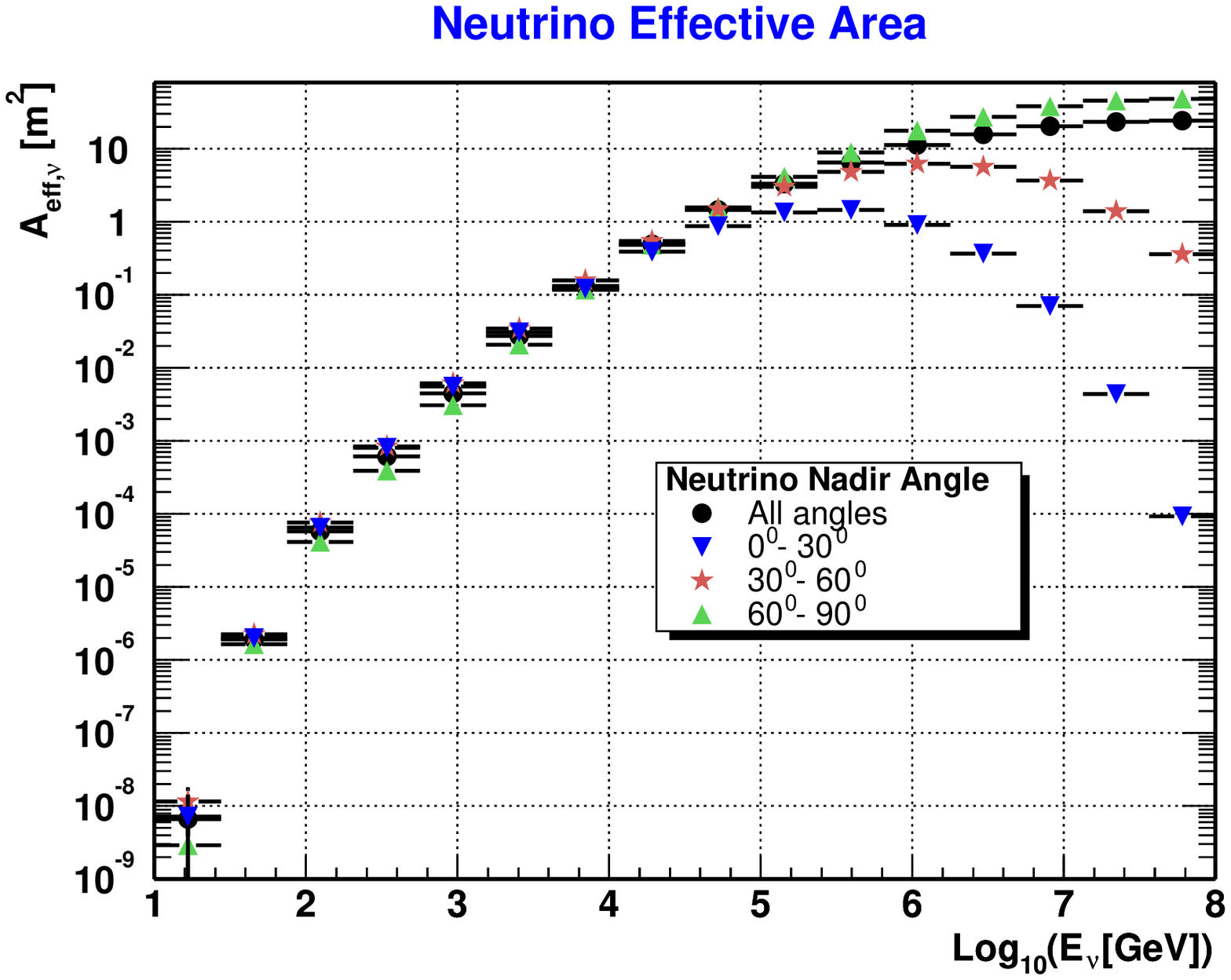,width=7cm}\hskip -0cm\epsfig{figure=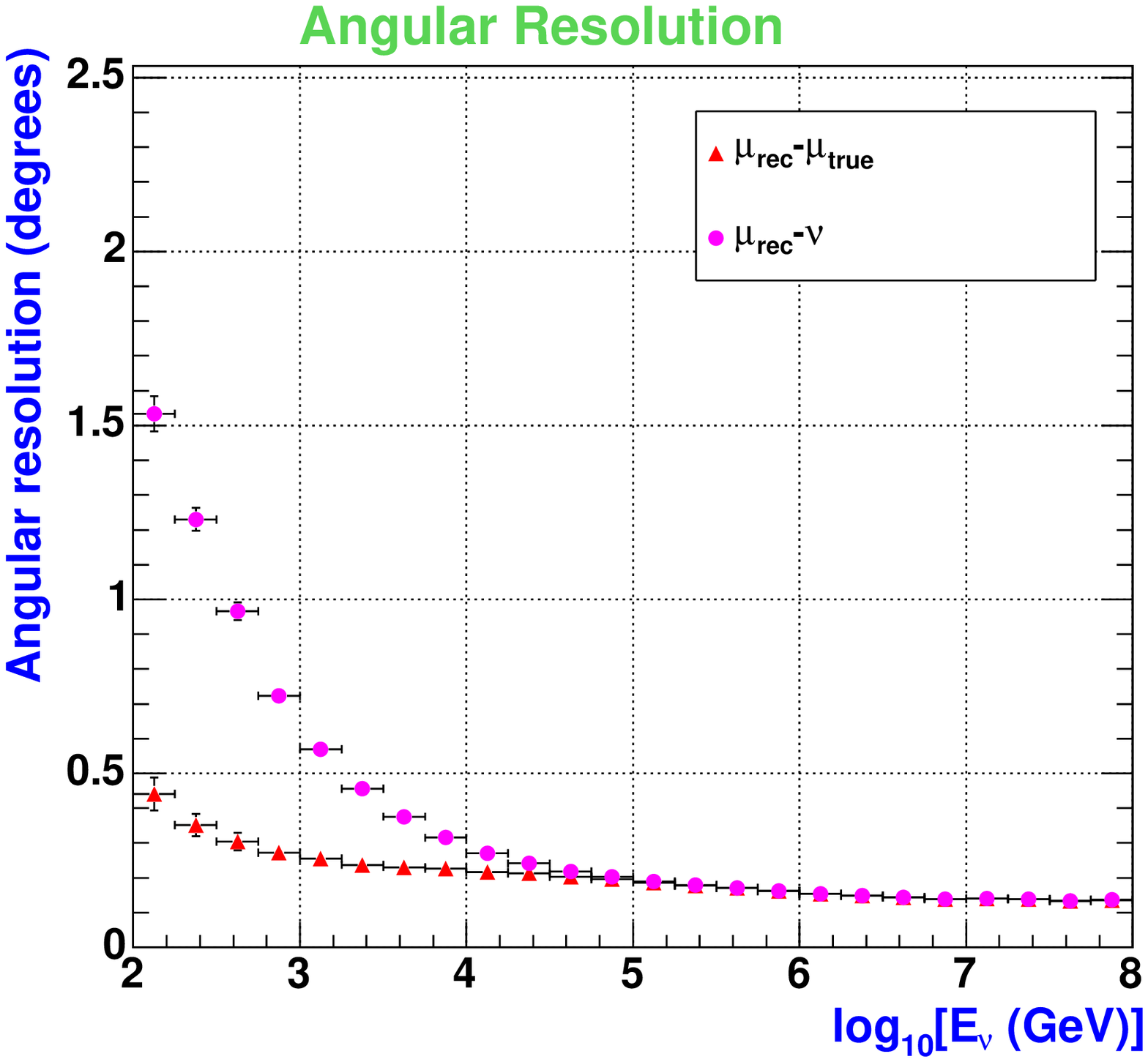,width=6cm}}

\caption{Effective area vs neutrino energy for various zenith angles (left) and expected angular resolution both for muons and neutrinos (right).} 
   \label{antares_perf}
\end{figure}

The {\it effective area} is defined as the ratio of detected neutrinos per unit of time over the incident neutrino flux. 
For angles around the vertical, as can be seen in Figure~\ref{antares_perf}, it reaches its maximum for an energy $E_{max} \simeq 10^5 eV$, 
and then decreases because of the shadowing of the Earth. 

The {\it angular resolution} is constant and equal to $0.2^{\circ}$ (48 arcsec)
for energies higher than 10 TeV as shown in Figure \ref{antares_perf}; below this energy, the resolution is dominated by kinematics.
Above 10 TeV \anta~will really be able to pinpoint a source in the sky, with the same resolution as the actual satellites dedicated to
one of the most promising sources for high energy neutrinos, namely Gamma-Ray Bursters.

Due to its location, \anta~will cover a 3.5$\pi~sr$ fraction of the sky.
{\sc Amanda}, the largest currently operational experiment, and soon {\sc IceCube}, located at the South Pole, will only cover 2$\pi~sr$, but 
with the same exposition during the day. Furthermore, \anta~will be able to observe the Galactic Centre, where {\sc Integral} 
has recently observed a great deal of new point-like gamma sources. The two detectors will nevertheless 
be complementary with an instantaneous overlap of 0.5$\pi~sr$.

\subsection{A step forward in Multi-Messenger Astronomy}\label{microq}

With the forthcoming operation of {\sc Hess} or {\sc Glast} for $\gamma$, {\sc Auger} for UHE cosmics rays, {\sc Virgo} for GW and \anta~for high energy $\nu$,
the years to come will bring a harvest of new information related to the most energetic and powerful phenomena in the universe. 

Good candidates for such a multi-messenger approach are microquasars, the galactic equivalents of quasars. More than 
10 of these objects have been observed
in our galaxy in all electromagnetic domains, and some models show that \anta~could detect up to tens of events per year from 
some of them \cite{microq}.
Furthermore, if the ultra-relativistic plasma ``blob'' ejected by microquasars is compact enough, efficient GW emission could take place \cite{microqgw}, 
thus making the coincident neutrino emission easier to detect by \anta.

\section{Deployments and Prototype Lines}

\begin{figure}[t!]
\begin{center}
\vskip 0.075cm
\epsfig{figure=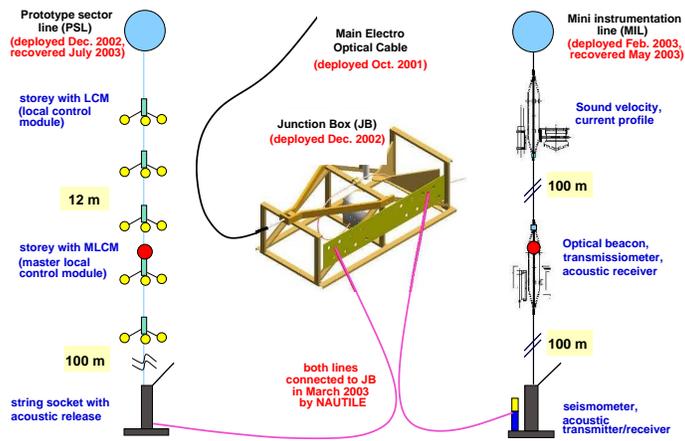,width=9cm}
\end{center}
\caption{Description of the prototype lines connected in 2003.} 
   \label{antares_scheme}
\end{figure}

Before launching the mass production of the lines, two prototypes of the 
final lines were built and deployed. Figure \ref{antares_scheme} illustrates the structure of the prototypes, together with the dates of previous deployments.

The deployments, connections and recovery of the lines were successful, and most of the components worked properly. One major problem showed up, though.
Because of a broken optical fibre in the cable of the Prototype Sector Line (PSL), the clock signal only reached the bottom of the String Control Module, thus 
preventing any coincidence measurements between storeys. The accuray in time calibration thus only reached $\sim 1ms$.

In about 100 days of running time, a large amount of data 
has nonetheless been recorded, 
and background light counting rates were extracted.

\subsubsection{Background Light \& Flow-induced Bioluminescence}

\begin{figure}[t!]  
\vskip -0.5cm
\centerline{\hskip -0cm \epsfig{figure=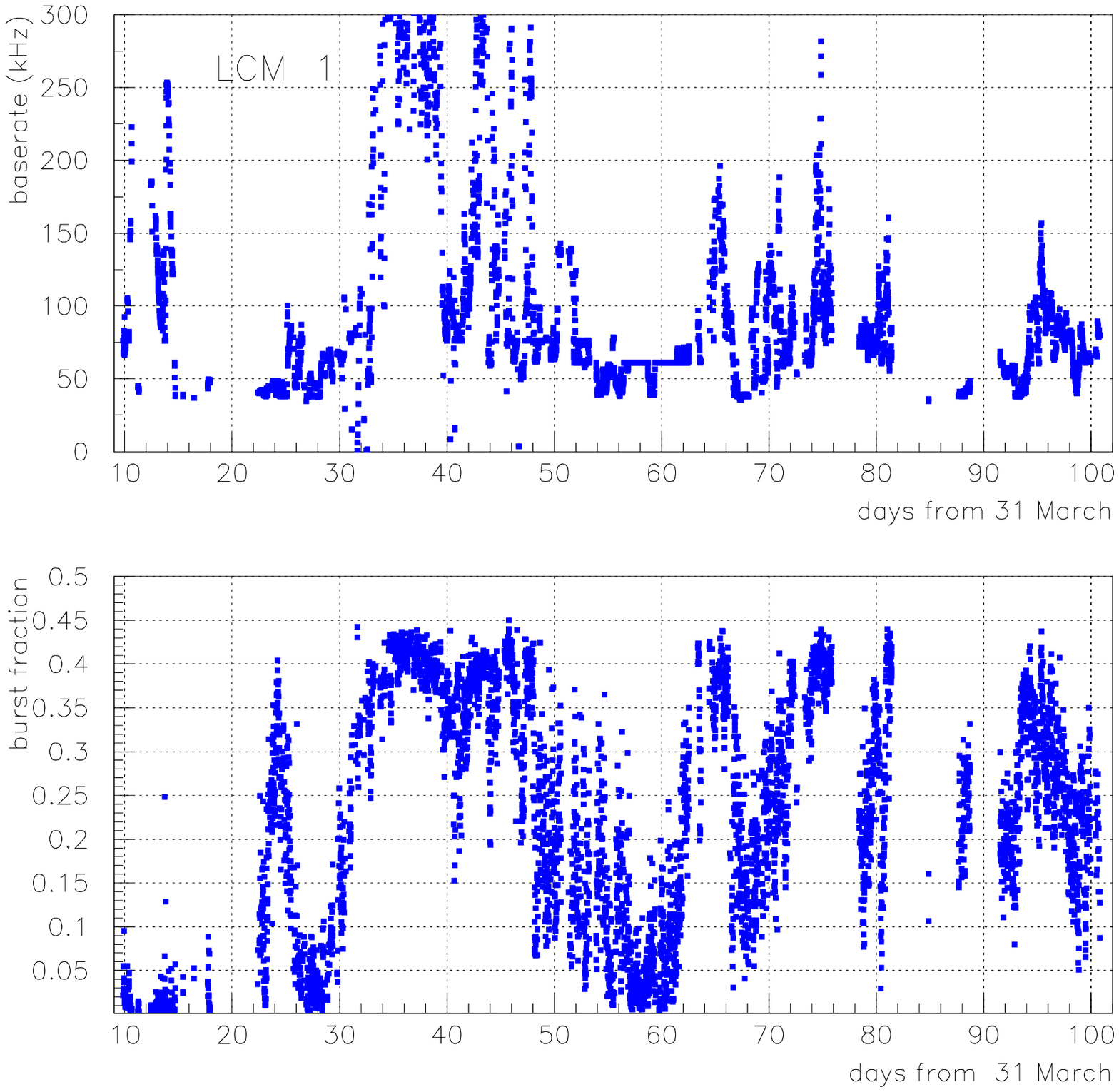,width=6cm}\hskip -0cm\epsfig{figure=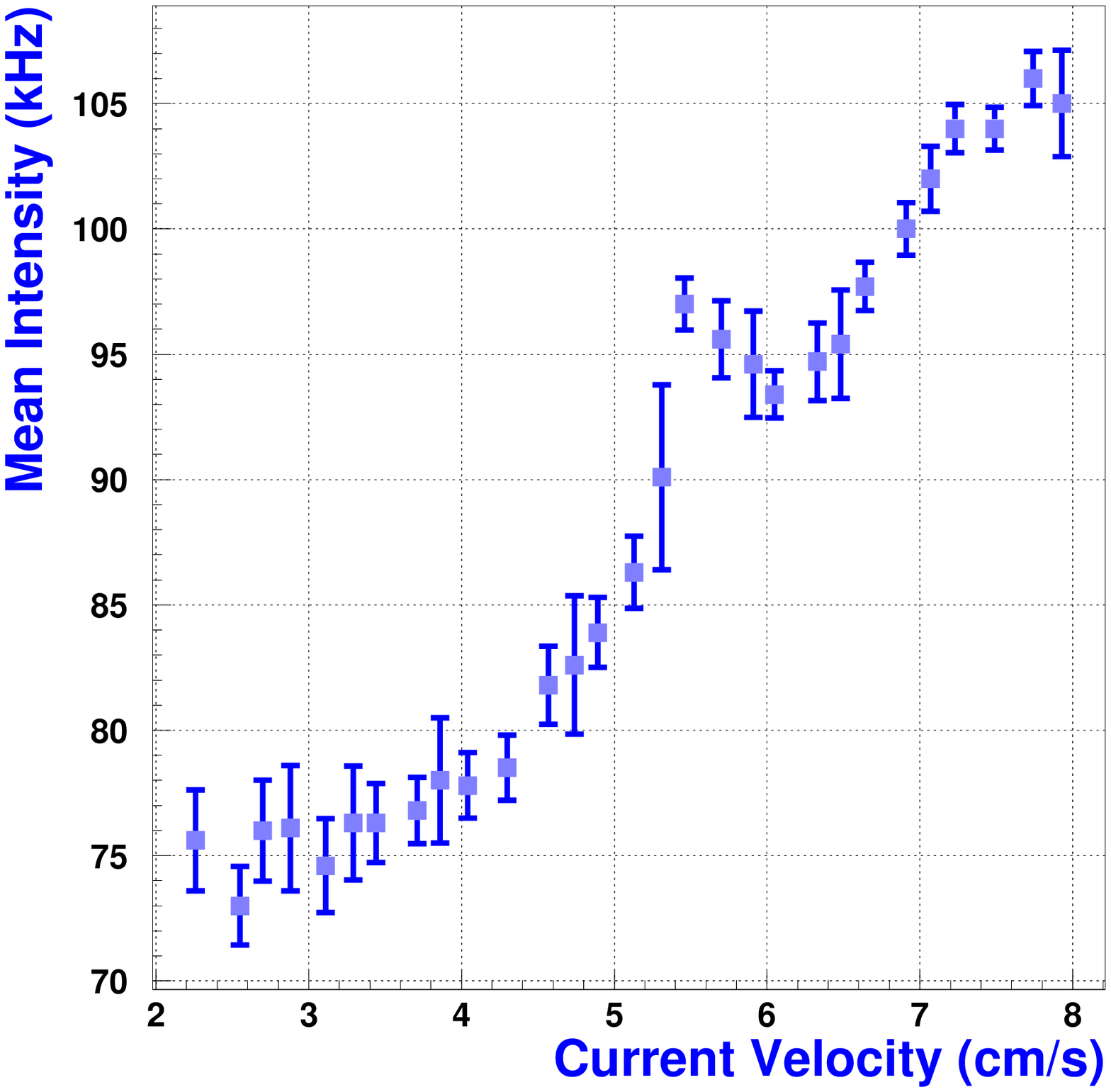,width=6cm}}
\caption{Variations of background light burst/continuous components (PSL data, left). Correlation between mean counting rate 
and sea current (right).} 
   \label{antares_biolum}
\end{figure}

The counting rates as recorded by the PMTs shows large and short-lived peaks, due to bioluminescent organisms (plankton), over a continuous
base-line coming both from $\beta$-decay of $^{40}$K and bacteria, as previously observed~\cite{water}. Both the fraction of bursts and 
the continuous level are subject to great variations (see Figure \ref{antares_biolum}, left plot, for an example).

Correlations with sea currents for the mean counting rates showed behaviours which could be the signature of flow-induced bioluminescence 
in the turbulences/boundary layers around the detector itself (see \cite{biolum} and references therein). Figure \ref{antares_biolum} (right plot) shows the mean intensity (in kHz)
as a function of sea current velocity.

\section{Milestones for \anta~and {\sc km3Net}}

The past years have seen the installation/operation of some of the key components in \anta.
Next year, the PSL will be redeployed with a new electro-mechanical cable, together with an improved instrumentation line, and the mass production
for all the lines will be launched. The 
\anta~neutrino telescope is now scheduled to be operational by 2007, but physics studies 
will begin before its completion. 

\vskip 0.25cm

\anta~must be seen as the first stage toward a km$^3$-scale telescope, for which European institutes involved in current neutrino
astronomy pro-jects ({\sc Antares}, {\sc Nemo} and {\sc Nestor}) are already collaborating. This network, {\sc km3NeT} \cite{km3}, will give birth to
a telescope with which neutrinos will be as common a messenger as gamma-rays are now.

%

%

\begin{thebibliography}{99}
\bibitem{anta} \anta~Collaboration, {\em astro-ph/9707136} \& {\em astro-ph/9907432} 
\bibitem{microq} G. F. Burgio, {\em Detecting High-Energy $\nu$ from microquasars with \anta}, proceedings of the V$^{th}$ microquasars workshop,
 Beijing (2004), {\em astro-ph/0407339}
\bibitem{microqgw} E. B. Segalis {\it et al.}, {\em Emission of gravitational radiation from ultra-relativistic sources}, Phys. Rev. 
{\bf D64}, 064018 (2001)
\bibitem{water} \anta~Collaboration,{\em~Background light in potential sites for the \anta~undersea neutrino telescope}, 
Astropart. Phys. {\bf 13} 127-136 (2000)
\bibitem{biolum} A.-S. Cussatlegras \& P. Le Gal, {\em~Bioluminescence of the dinoflagellate {\it Pyrocystis noctiluca} induced 
by laminar and turbulent Couette flow}, Journal of Experimental Marine Biology and Ecology, in Press (2004) (available online 
at {\it http://www.sciencedirect.com})
\bibitem{km3} {\em http://www.km3net.org}
\end{thebibliography}
\end{document}